# Consensus formation Online using Sociophysics method

Yasuko Kawahata[1], Akira Ishii[2]

[1] Gunma University, Faculty of Social and Information Studies, 4-2 Aramaki-machi, Maebashi, Gunma, 371-8510, Japan, kawahata@gunma-u.ac.jp

[2] Department of Applied Mathematics and Physics, Tottori University, Koyama, Tottori, 680-8552, Japan, ishii@tottori-u.ac.jp

**Abstract.** Consensus formation and difference of opinion have long been the subject of research. However, relevant laws and systems within society are being updated to reflect the changes in information networks. Online environment has come to fulfill a major role as a real and concrete place of opposing opinions and consensus formation. In the future, quantitative findings on consensus formation, and findings on relevant trends, must be summarized, and quantitative research related to trends likely to give rise to social and economic risk is required. Thus, the potential for comparing research related to consensus formation using actual data and an approach using a mathematical model was first investigated.

## 1 Introduction

Consensus formation and difference of opinion have long been the subject of research [1-6]. However, relevant laws and systems within society are being updated to reflect the changes in information networks [7]. After a notable global diffusion of public network devices from the launch of Microsoft Windows 95 in 1995, there was a substantial increase in the number of opportunities for decision-making and consensus formation transcending time and space. In other words, the online environment has come to fulfill a major role as a real and concrete place of opposing opinions and consensus formation. We have entered an age in which it is possible for a large volume of data on our online behavior to be accumulated, in the form of public network communication logs.

In particular, the advent of major companies (including the so-called "GAFA" or "FANG") with telecommunications platforms allowing the acquisition of life log data indicates that our everyday data, whether consciously or unconsciously, are sometimes treated as a form of consensus formation [8].

With these social environment systems, the online pseudo society produced by public networks has come to be visualized in the same way as the natural universe. Thus, since public networks became ubiquitous, there has been rapid growth in research in the field of web science, which seeks to explain the mechanisms of social phenomena that were previously difficult to visualize or quantify [9]. Research and development that applies these disciplines and results in social practice has also been proposed as a field of data science and is gradually progressing. Many worrying factors, such as the rise in fake news and the risk to society of excessive online opinion clashes impacting directly on ordinary life, are now being referred to in the news media and in literature. To provide a recent example, it has been possible to view the gilets jaunes demonstrations in Paris in France in real time on YouTube [10]. In the future, quantitative findings on consensus formation, and findings on relevant trends, must be summarized, and quantitative research related to trends likely to give rise to social and economic risk is required. Thus, the potential for comparing research related to consensus formation using actual data and an approach using a mathematical model was first investigated.

## 2  Modeling opinion dynamics

Our model is based on the original bounded confidence model of Hegselmann-Krause[11]. For a fixed agent, say $i$, where $1 \leq i \leq N$, we denote the agent's opinion at time t by $I_i(t)$. As shown in Fig.1, person $i$ can be affected by surrounding people. According to Hegselmann-Krause[11], opinion formation of agent $i$ can be described as follows.

$$I_i(t+1) = \sum_{j=1}^{N} D_{ij} I_j(t) \qquad (1)$$

This can be written in the following form.

$$\Delta I_i(t) = \sum_{j=1}^{N} D_{ij} I_j(t) \Delta t \qquad (2)$$

where it is assumed that $D_{ij} \geq 0$ for all $i, j$ in the model of Hegselmann-Krause. Based on this definition, $D_{ij} = 0$ means that the opinion of agent $i$ is not affected by the opinion of agent $j$. In this theory, it is expected implicitly that the final goal of the negotiation among people is the formation of consensus.

However, in the real society in the world, the formation of consensus among people is sometimes very difficult. We can find many such examples in the international politics like the former East-West cold war. Even in domestic problems, the opinions between people pursuing economic development and people claiming nature conservation are not compatible and agreement is difficult. Since it is not possible to define the payoff matrix for such serious political conflict, application of game theory may be difficult. Thus, in order to deal with problems that are difficult to form consensus among these people, it is necessary to include the lack of trust between people in our opinion dynamics theory. Here, as a result of exchanging opinions, consider the possibility that the opinions of two people with different opinions change move in different directions. Let's us think about consider the distribution of opinions with in the positive and negative directions of a one-dimensional axis.

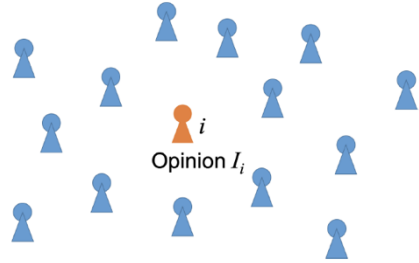

**Fig. 1.** Schematic illustration of opinion dynamics think

In this case, the value range of $I_i(t)$ is
$$-\infty \leq I_i(t) \leq +\infty.$$
We modify the meaning of the coefficient $D_{ij}$ as the coefficient of trust. We assume here that $D_{ij} > 0$ if there is a trust relationship between the two persons, and $D_{ij} < 0$ if there is distrust relationship between the two persons. In contrast to our previous theory[35], we consider here that that people disregard the opinion far removed from their opinions without agreeing or repelling. Also, opinions that are very close to himself/herself will not be particularly affected. To include the two effects, we use the following function instead of $D_{ij}I_j(t)$ as follows,

$$D_{ij}\Phi(I_i, I_j)(I_j(t) - I_i(t)) \qquad (3)$$

where

$$\Phi(I_i, I_j) = \frac{1}{1 + exp(\beta(|I_i - I_j| - b))} \qquad (4)$$

This function can work as a smooth cut-off function at $|I_i - I_j| = b$. Moreover, because of the factor $I_j(t) - I_i(t)$, the opinion $I_i(t)$ is not affected by the opinion $I_j(t)$ if the opinion $I_j(t)$ is almost same as the opinion $I_i(t)$. Influences of mass media and government statements can not be ignored in the formation of public opinion. Such mass media effect can also work even for negotiations of small size group. Since formula of our theory above is similar to the model of hit phenomena[12] where the popularity of certain topic is analyzed using the sociophysics model, we introduce here the effects of mass media similar to the way of ref[13]. Let $A(t)$ be the pressure at time $t$ from the outside and denote the reaction difference for each agent is denoted by the coefficient $c_i$. The coefficient $c_i$ can have different values for each person and $c_i$ can be positive or negative. If the coefficient $c_i$ is positive, the person $i$ moves the opinion toward the direction of the mass media. On the contrary, if the coefficient $c_i$ is negative, the opinion of the person change against the mass media direction.
Therefore, including such mass media effects, the change in opinion of the
agent can be expressed as follows.

$$\Delta I_i(t) = c_i A(t) \Delta t + \sum^N D_{ij} \Phi(I_i(t), I_j(t))(I_j(t) - I_i(t)) \Delta t \qquad (5)$$

$$\Delta I_i(t) = -\alpha I_i(t) \Delta t + c_i A(t) \Delta t + \sum_{j=1}^N D_{ij} \Phi(I_i(t), I_j(t))(I_j(t) - I_i(t)) \Delta t \qquad (6)$$

We assume here that $D_{ij}$ and $D_{ji}$ are independent. Usually, $D_{ij}$ is an asymmetric matrix; $D_{ij} \neq D_{ji}$. Moreover, $D_{ij}$ and $D_{ji}$ can have different signs.
Long-term behavior requires attenuation, which means that topics will be
forgotten over time. Here we introduce exponential attenuation. The expression is as
follows.

## 3  Method of data acquisition

**3.1 Data**
  In this section, we discuss which data best capture people's reaction to information transmitted by media in society. Currently, there are also various studies related to natural language processing owing to i) new full provision of data based on dictionary data and learning data and ii) expansion of classification methods in machine learning [14-16]. In short, as it is now possible to judge the strength or weakness of opinion based on text collected in natural speech processing, it is also possible now to measure the distribution of opinion within society not in binary terms but as continuous distribution from positive to negative and neutral comments. For this study, given the constraints on data acquisition from various media across the world, YouTube was selected. Various online social networks allow streaming in real time, but media companies that own the world's leading TV and press outlets also publish snippets of their news reports on YouTube. In addition, a characteristic of YouTube is that it allows real-time transmission. In addition, it has been hypothesized that it is possible to discuss the distribution of opinion in terms of i) transmission content for the news media in each country; ii) difference in reporting of the same news content; iii) difference in users who make comments. In addition, in this report, in order to identify the weakness and strength of opinion under the same conditions, analysis was limited to news videos on channels based on Anglophone dictionary data. Regarding YouTube, although there are cases related to language characteristics, video content is globally transmitted, and analysis was performed based on the hypothesis of similarity to transmission on TV in terms of comparison being expected related to the difference between diff erent media types and in terms of there being elements where, even for the same report content, the nature of reception differs according to the method of reporting.

## 3.2 Calculate

Regarding the acquisition target channels, in this case, the main channel was CNN. The period of comment acquisition, for Case A, was from 2012/11/15 to 2018/11/30 (5,020 comments) and, for Case B, was from 2015/6/5 to 2018/12/2 (3,163comments). In this study, the comments related to videos with the most views were collected since the launch of the CNN channel as a trial and, for each comment, the negative, positive, and neutral scores were totaled and processed so that they became 1. The scores on all three axes were defined in the range of [-1.0, -0.75, -0.5, -0.25, 0, 0.25, 0.5, 0.75, 1.0]. In this case, the process was abbreviated as -1, 0, and 1,which are bias scores when distribution is considered.

To analyze the natural language input, the Natural Language Toolkit[17] was used. Utilizing vader analyze, feature value extraction was performed for each separate comment in each video and, finally, the distribution of the negative, positive, and neutral scores for all comments for each video was output onto the three axes. In addition, when the negative, positive, and neutral scores for each axis were selected in integrated distribution, the method adopted was adding 1 for a positive opinion, -1 for a negative opinion, and zero for a neutral opinion, with conversion to a scale of 0 to 2.

## 4    Measurement Case

CNN's video ranking by number of views is biased toward animal videos and amusing videos. In this study, cases where an apparent negative or positive bias could be expected, such as news items with political ideas and disasters, were excluded. The focus was on comment-opinion distribution in animal videos with a high number of views.

### 4.1 Case A "Skydiving cats cause uproar"

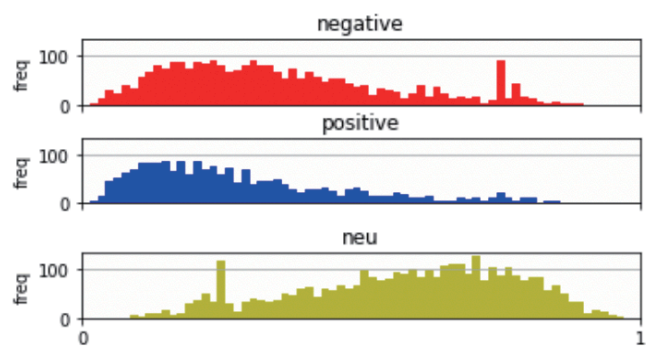

**Fig. 2.** Distribution of negative, positive, and neutral comments in Case A (acquisition period 2012/11/15 to 2018/11/30; score range in each case: 0 to1)

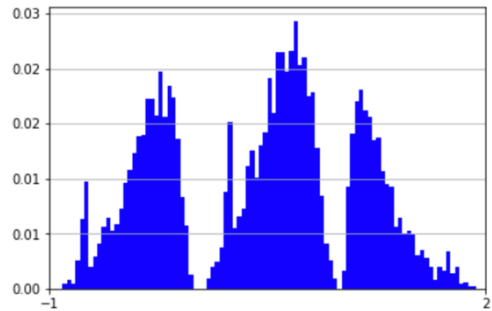

**Fig. 3.** Distribution when the negative, positive, and neutral comments on Case A are integrated (score range: 0 to 2) (acquisition period 2/11/15 to 2018/11/30)

**4.2 Case A "Skydiving cats cause uproar"**

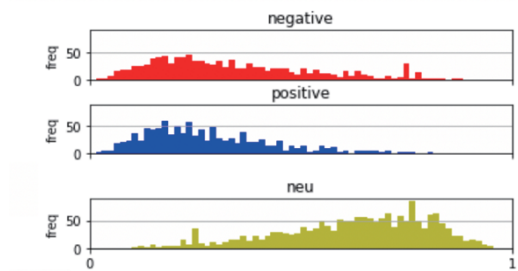

**Fig. 4.** Distribution of negative, positive, and neutral comments in Case B (acquisition period 2015/6/5 to 2018/12/2; score range in each case: 0 to1)

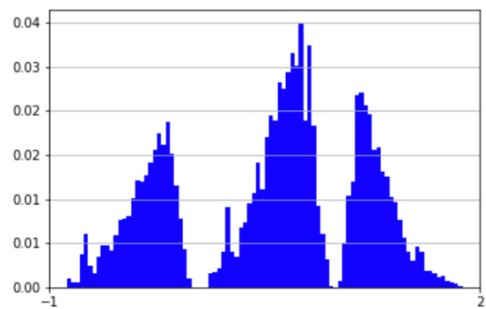

**Fig. 5.** Distribution of negative, positive, and neutral comments in Case B (acquisition period 2015/6/5 to 2018/12/2)

Case A and Case B are both animal videos and, while negative mentions are notable at the extreme, neutral opinion is somewhat beyond 1, and it is inferred that the tendency is to take a negative outlook. This example can be surmised to be one where there is a division between critical opinion, from the viewpoint of animal protection, and those who find the videos funny.

# 5 Discussion

The calculated opinion distribution is flat. In addition, as the distribution of neutral opinion was measured, the trend differs from simulation results for calculations for cases in which opinion is polarized. A case is assumed in which opinion distribution according to simulation results separates into a number of clusters. This shows that it is necessary to improve further the reflection within the theory of interaction between people. This result suggests a scope for both gaining understanding related to discrepancy from actual measurement and for improvement in the model. In addition, notably, there are elements that allow us to surmise trends in actual society and similarity relationship trends from the viewpoint of sociology. It is possible to measure the distribution of opinion regarding a problem even when, for example, the positive-negative balance has been destroyed by, for example, reports of a sudden event causing the distribution to change. By performing simulations according to new opinion dynamics theory, our objective is to attain a better understanding of quantitative trends in the discussion related to the formation of radical public opinion or a social crisis that occurs online where decision-making and opinion exchange in the formation of public opinion can occur.